# Observation of Fermi arc spin texture in TaAs

B. Q. Lv[1,2,*], S. Muff[2,3,*], T. Qian[1,*], Z. D. Song[1,*], S. M. Nie[1], N. Xu[2], P. Richard[1,4], C. E. Matt[2], N. C. Plumb[2], L. X. Zhao[1], G. F. Chen[1,4], Z. Fang[1,4], X. Dai[1,4], J. H. Dil[2,3], J. Mesot[2,3,5], M. Shi[2,§], H. M. Weng[1,4,§], and H. Ding[1,4,§]

[1] *Beijing National Laboratory for Condensed Matter Physics and Institute of Physics, Chinese Academy of Sciences, Beijing 100190, China*

[2] *Swiss Light Source, Paul Scherrer Institute, CH-5232 Villigen PSI, Switzerland*

[3] *Institute of Condensed Matter Physics, École Polytechnique Fédérale de Lausanne, CH-1015 Lausanne, Switzerland*

[4] *Collaborative Innovation Center of Quantum Matter, Beijing, China*

[5] *Laboratory for Solid State Physics, ETH Zürich, CH-8093 Zürich, Switzerland*

\* These authors contributed equally to this work.
§ E-mail: ming.shi@psi.ch, hmweng@iphy.ac.cn, dingh@iphy.ac.cn

Abstract

We have investigated the spin texture of surface Fermi arcs in the recently discovered Weyl semimetal TaAs using spin- and angle-resolved photoemission spectroscopy. The experimental results demonstrate that the Fermi arcs are spin-polarized. The measured spin texture fulfills the requirement of mirror and time reversal symmetries and is well reproduced by our first-principles calculations, which gives strong evidence for the topologically nontrivial Weyl semimetal state in TaAs. The consistency between the experimental and calculated results further confirms the distribution of chirality of the Weyl nodes determined by first-principles calculations.

While the Weyl fermion as a fundamental particle remains evasive after its prediction 86 years ago [1], the low-energy Weyl fermion quasiparticle has been recently predicted theoretically [2, 3] and observed experimentally in TaAs [4-7] and related materials [8-12]. These so-called Weyl semimetals (WSMs) possess pairs of Weyl nodes with opposite chirality in the bulk. A Weyl node is a singularity point of Berry curvature, which can be viewed as a magnetic monopole in momentum space [13-15]. Direct detection of the sign of chiral charge (or magnetic polarity) of Weyl node is fundamentally important, but highly challenging.

A direct consequence of Weyl nodes is the emergence of surface Fermi arcs [16, 17] connecting the projections of Weyl nodes with opposite chiral charge on the surface. The Fermi arcs are the traces at the Fermi energy ($E_F$) of the chiral edge states, which are protected by the topologically nontrivial phase arising from the separation of Weyl nodes with opposite chiral charge. Since the edge state is spin non-degenerate, the Fermi arcs naturally possess a momentum distribution of spin, namely spin texture (as shown schematically in Fig. 1(a)), which can be directly detected by spin- and angle-resolved photoemission spectroscopy (SARPES).

In this work, we have investigated the spin texture of the surface Fermi arcs in TaAs using SARPES. Our experimental results demonstrate that the Fermi arcs are spin-polarized and the measured spin texture is consistent with our first-principles calculations, which further proves the existence of topologically nontrivial WSM states in TaAs. More importantly, the chirality of the Weyl nodes beneath their projections connected by the Fermi arcs can be convincingly inferred through comparison between calculations and measurements.

High-quality TaAs single crystals were grown by the chemical vapor transport method, as described in Ref. [18]. Conventional ARPES measurements were performed at the Dreamline beamline at the Shanghai Synchrotron Radiation Facility (SSRF) with a Scienta D80 analyzer, and at the Surface/Interface Spectroscopy (SIS) beamline at the Swiss Light Source (SLS) with a Scienta R4000 analyzer. SARPES measurements were performed at the COPHEE endstation at the SIS beamline at the SLS with an Omicron EA 125 hemispherical energy analyzer equipped with two orthogonally mounted Mott detectors [19]. Figure 1(b) schematically illustrates the geometry of the SARPES measurements. The angular and energy resolutions in the spin-resolved mode were set to 1.5° and 60 meV, respectively. In the spin-integrated mode the resolutions of the COPHEE set-up were set to 0.5° and 20 meV. The spot size of the incident beam was 200×200 μm in the SARPES measurements. Further technical details of the data acquisition and analysis can be found in Ref. [20]. Samples with a typical size of ~1×1 mm$^2$ were cleaved *in situ* and measured at 20 K in a working vacuum better than 5×10$^{-11}$ Torr. We show in Figs. 1(c) and 1(d), respectively, the crystal structure of TaAs and the corresponding bulk as well as (001) surface Brillouin zones (BZs), along with the notations used for the high-symmetry points.

Figure 1(e) shows the overall surface state electronic structure on the (001) surface of TaAs recorded by conventional ARPES measurements. The surface state band dispersions exhibit a moderate anisotropy along the $\bar{\Gamma}-\bar{X}$ and $\bar{\Gamma}-\bar{Y}$ directions due to the breaking of the four-fold symmetry on the cleavage surface. Here we focus on the spin-polarization texture of the surface state bands near the middle points between $\bar{\Gamma}$ and $\bar{X}(\bar{Y})$ in the subsequent SARPES measurements as they are most clearly identified. As shown in Fig. 2(a), these bands form horseshoe-like Fermi surfaces (FSs) around the Weyl nodes W1.

We display in Figs. 2(b) and 2(c) the spin-integrated intensity maps of the near-$E_F$ bands along $\bar{\Gamma}-\bar{X}$ (*C*1) and $\bar{\Gamma}-\bar{Y}$ (*C*3), respectively. Although the momentum and energy resolutions in the COPHEE set-up are lower than in conventional ARPES systems, the bands are clearly identified, indicating the high quality of the cleaved sample surface. This enables us to determine the respective spin polarization of these bands. The bands *a*1 (*b*1) and *a*2 (*b*2) form the inner and outer FSs near $\bar{\Gamma}-\bar{X}$ ($\bar{\Gamma}-\bar{Y}$), respectively. In our previous work, the inner FSs were assigned as trivial FSs while the outer ones were assigned as nontrivial Fermi arcs, which connect the adjacent W1 Weyl nodes with opposite chiral charges [4].

We plot in Figs. 2(d)-2(i) the spin-up ($I^\uparrow$) and spin-down ($I^\downarrow$) momentum distribution curves (MDCs) as well as the corresponding spin polarization spectra along $\bar{\Gamma}-\bar{X}$ at +$k_x$ (*C*1) in the *x*, *y* and *z* directions, respectively. While no significant spin polarization is observed in the *x* and *z* directions within our experimental resolution, the spin polarization components are large in the *y* direction. As shown in Fig. 2(h), the bands *a*1 and *a*2 have negative and positive spin polarization components in the *y* directions, respectively. These results indicate that the spins on *a*1 and *a*2 at +$k_x$ are polarized in the $-\hat{y}$ and $+\hat{y}$ directions, respectively. The spin polarization spectra in Figs. 2(j)-2(l) indicate that the spins on *a*1 and *a*2 at -$k_x$ (*C*2) are polarized in the $+\hat{y}$ and $-\hat{y}$ directions, respectively. These spins are thus reverse compared to those at +$k_x$. The measured spin texture on *a*1 and *a*2 at both +$k_x$ and -$k_x$ fulfills the time-reversal symmetry with respect to the time-reversal invariant points $\bar{\Gamma}$ and $\bar{X}$.

It has been shown that a spin polarization signal can appear in the photoemission process from states that possess no net spin polarization [21-24]. In contrast to the intrinsic spin signal from the spin-polarized initial states, the non-intrinsic spin signal caused by the photoemission effect depends on the incident photon energy and polarization. To check whether the spin signal is intrinsic, we performed SARPES measurements along *C*1 with different incident photon energies and polarizations. As shown in Figs. 2(h), 2(m), and 2(n), the consistent spin texture obtained with different photon energies and polarizations supports that the observed spin polarizations reflect the intrinsic spin structure of the initial states. Moreover, the data on *a*1 and *a*2 at +$k_x$ and -$k_x$ show that the spin texture follows the expected symmetry operation, providing further evidence of its intrinsic nature. Finally, as discussed below, the measured spin polarizations match the calculations remarkably well. All the results allow us to conclude that the measured spin polarizations are related to an intrinsic origin.

To determine the spin polarizations of the bands along $\bar{\Gamma}-\bar{Y}$, we have carried out SARPES measurements along *C*3. Similar to the observation along *C*1, the bands *b*1 and *b*2 along *C*3 are spin polarized, but now in the *x* direction [Fig. 2(o)]. We summarize the spin texture along $\bar{\Gamma}-\bar{X}$ and $\bar{\Gamma}-\bar{Y}$ in Fig. 2(a). While the spin polarizations on *b*1 and *b*2 were measured at +$k_y$, those at -$k_y$ are inferred based on the time-reversal symmetry with respect to the time-reversal invariant point $\bar{\Gamma}$.

After determining the spin texture along $\bar{\Gamma}-\bar{X}$ and $\bar{\Gamma}-\bar{Y}$, we focus on the spin texture on *b*1 and *b*2. Figures 3(c)-3(h) plot the spin polarization spectra along *C*4 and *C*5 in the *x*, *y* and *z* directions, respectively. As $\bar{\Gamma}-\bar{Y}$ is the projection in the (001) surface BZ of the mirror plane $\Gamma-\Sigma-S-Z$ in the bulk BZ [see Fig. 1(d)], the spin texture on *b*1 and *b*2 should fulfill the mirror symmetry with respect to $\bar{\Gamma}-\bar{Y}$. Upon moving away from $\bar{\Gamma}-\bar{Y}$, the spin polarizations would rotate clockwise on one side but counter-clockwise on the other side. As the spin polarizations on *b*1 and *b*2 along $\bar{\Gamma}-\bar{Y}$ are in the *x* direction, the mirror symmetry operation imposes to the spin polarizations at symmetric locations with respect to $\bar{\Gamma}-\bar{Y}$ to have equal $P_x$

(perpendicular to the mirror plane) signals but opposite $P_y$ and $P_z$ (parallel to the mirror plane) signals. The measured spin polarizations along C4 indeed fulfill the mirror symmetry while the four expected peaks along C5 are not adequately resolved. However, the related FSs are clearly resolved on the same cleaved surface measured in the spin-integrated mode, as shown in Fig. 3(a). The momentum resolutions in the spin-integrated and spin-resolved modes are 0.03 and 0.08 $\pi/a$, respectively. The unresolved MDC peaks along C5 in the spin-resolved mode could be attributed to the low momentum resolution, which is larger than the distance between two adjacent peaks.

The measured spin texture of surface states is consistent with the calculated one plotted in Fig. 3(b). This remarkable consistency builds further confidence in the WSM ground state of TaAs. The theoretical spin texture satisfies the necessary symmetrical constraints, such as the symmetries with respect to the mirror planes along $\overline{\Gamma}-\overline{Y}$ and $\overline{\Gamma}-\overline{X}$, respectively. It is known that the four-fold rotational symmetry with respect to the z-axis is broken at the termination of the (001) surface. However, the surface states roughly follow the four-fold symmetry as it is recovered inside the bulk. This can be noticed in the spin texture along the $\overline{\Gamma}-\overline{Y}$ and $\overline{\Gamma}-\overline{X}$ directions in Fig. 3(b).

For a specific material, the spin texture of surface states and the distribution of chirality of Weyl nodes are both fixed in theoretical calculations. Although there is neither a universal nor a generic relationship between them, the chirality of a Weyl node can be convincingly inferred from the consistency between the experimental and calculated spin textures of the surface states. By using the consistent coordinate system and crystal structure in both experimental setting and theoretical calculation, we can identify the positive (negative) chirality for Weyl node $+\vec{k}\cdot\vec{\sigma}$ ($-\vec{k}\cdot\vec{\sigma}$) in Fig. 3(b) according to the distribution of calculated Berry curvature for the lower band of Weyl cone [2, 15].

Finally we draw attention to the difference between the spin texture of surface Fermi arc of a WSM and those of surface states of a 3D topological insulator (TI) and of a Dirac semimetal (DSM). The surface state of a TI has a characteristic spin texture illustrated in Fig. 4(a), which is highly symmetrical as the Fermi level is close enough to the Dirac point at time-reversal invariant momenta. For a DSM due to band inversion, for example $Na_3Bi$ [25] and $Cd_3As_2$ [26], the spin texture of Fermi arcs is quite similar to that of surface states in 3D TIs, except at the touching point of two arcs, where the spin degeneracy recovers and the spin is ill-defined, as shown in Fig. 4(b). For a noncentrosymmetric WSM satisfying time reversal symmetry, the Weyl points are at arbitrary momenta. The Fermi arc and its spin texture have no specific symmetry. In TaAs, only the mirror symmetry and time reversal symmetry constrain the spin texture, as demonstrated by our SARPES measurements and first-principles calculations.

In summary, our SARPES measurements have demonstrated that the Fermi arcs are spin-polarized. The measured spin texture is consistent with our first-principles calculations, which builds further confidence in the topologically nontrivial WSM states in TaAs. The chirality of the Weyl nodes is convincingly inferred through comparison between calculations and measurements. This will stimulate successive experimental efforts in further studying and even fine-tuning the chirality of Weyl nodes inside the bulk of WSMs.

This work was supported by the Ministry of Science and Technology of China (No. 2013CB921700, No. 2015CB921300, No. 2011CBA00108, and No. 2011CBA001000), the National Natural Science Foundation of China (No. 11474340, No. 11422428, No. 11274362, and No. 11234014), the Chinese Academy of Sciences (No. XDB07000000), the Sino-Swiss Science and Technology Cooperation (No. IZLCZ2138954), and the Swiss National Science Foundation (No. 200021-137783 and No. PP00P2_144742/1).

Note: After the submission of our manuscript, we noticed that a very recent publication offers calculation of spin polarization of the Fermi arcs [27], which is fully consistent with our calculations.

**Reference**

[1]  H. Weyl, Z. Phys. **56**, 330 (1929).

[2]  H. Weng, C. Fang, Z. Fang, B. A. Bernevig, and X. Dai, Phys. Rev. X **5**, 011029 (2015).

[3]  S.-M. Huang, S.-Y. Xu, I. Belopolski, C.-C. Lee, G. Chang, B. Wang, N. Alidoust, G. Bian, M. Neupane, C. Zhang, S. Jia, A. Bansil, H. Lin, and M. Z. Hasan, Nat. Commun. **6**, 7373 (2015).

[4]  B. Q. Lv, H. M. Weng, B. B. Fu, X. P. Wang, H. Miao, J. Ma, P. Richard, X. C. Huang, L. X. Zhao, G. F. Chen, Z. Fang, X. Dai, T. Qian, and H. Ding, Phys. Rev. X **5**, 031013 (2015)

[5]  S.-Y. Xu, I. Belopolski, N. Alidoust, M. Neupane, G. Bian, C. Zhang, R. Sankar, G. Chang, Z. Yuan, C.-C. Lee, S.-M. Huang, H. Zheng, J. Ma, D. S. Sanchez, B. Wang, A. Bansil, F. Chou, P. P. Shibayev, H. Lin, S. Jia, and M. Z. Hasan, Science, (2015)

[6]  B. Q. Lv, N. Xu, H. M. Weng, J. Z. Ma, P. Richard, X. C. Huang, L. X. Zhao, G. F. Chen, C. E. Matt, F. Bisti, V. N. Strocov, J. Mesot, Z. Fang, X. Dai, T. Qian, M. Shi, and H. Ding, Nature Phys. **11**, 724 (2015).

[7]  L. X. Yang, Z. K. Liu, Y. Sun, H. Peng, H. F. Yang, T. Zhang, B. Zhou, Y. Zhang, Y. F. Guo, M. Rahn, D. Prabhakaran, Z. Hussain, S.-K. Mo, C. Felser, B. Yan, and Y. L. Chen, Nature Phys. **11**, 728 (2015).

[8]  N. Xu, H. M. Weng, B. Q. Lv, C. Matt, J. Park, F. Bisti, V. N. Strocov, D. Gawryluk, E. Pomjakushina, K. Conder, N. C. Plumb, M. Radovic, G. Autès, O. V. Yazyev, Z. Fang, X. Dai, G. Aeppli, T. Qian, J. Mesot, H. Ding, and M. Shi, arXiv:1507.03983.

[9]  S.-Y. Xu, N. Alidoust, I. Belopolski, Z. Yuan, G. Bian, T.-R. Chang, H. Zheng, V. N. Strocov, D. S. Sanchez, G. Chang, C. Zhang, D. Mou, Y. Wu, L. Huang, C.-C. Lee, S.-M. Huang, B. Wang, A. Bansil, H.-T. Jeng, T. Neupert, A. Kaminski, H. Lin, S. Jia, and M. Z. Hasan, Nature Phys. **11**, 748 (2015).

[10] D. F. Xu, Y. P. Du, Z. Wang, Y. P. Li, X. H. Niu, Q. Yao, P. Dudin, Z.-A. Xu, X. G. Wan, and D. L. Feng, Chinese Phys. Lett. **32**, 107101 (2015).

[11] I. Belopolski, S.-Y. Xu, D. Sanchez, G. Chang, C. Guo, M. Neupane, H. Zheng, C.-C. Lee, S.-M. Huang, G. Bian, N. Alidoust, T.-R. Chang, B. Wang, X. Zhang, A. Bansil, H.-T. Jeng, H. Lin, S. Jia, and M. Z. Hasan, arXiv:1509.07465.

[12] S. Souma, Z. Wang, H. Kotaka, T. Sato, K. Nakayama, Y. Tanaka, H. Kimizuka, T. Takahashi, K. Yamauchi, T. Oguchi, K. Segawa, and Y. Ando, arXiv:1510.01503.

[13] Z. Fang, N. Nagaosa, K. S. Takahashi, A. Asamitsu, R. Mathieu, T. Ogasawara, H. Yamada, M. Kawasaki, Y.Tokura, and K. Terakura, Science **302**, 92(2003).

[14] H. M. Weng, R. Yu, X. Hu, X. Dai and Z. Fang, Adv. Phys. **64**, 227 (2015).

[15] M. V. Berry, Proc. R. Soc. Lond. A **392**, 45 (1984).

[16] X. Wan, A. M. Turner, A. Vishwanath, and S. Y. Savrasov, Phys. Rev. B **83**, 205101 (2011).

[17] G. Xu, H. M. Weng, Z. Wang, X. Dai, and Z. Fang, Phys. Rev. Lett. **107**, 186806 (2011).

[18] X. Huang, L. Zhao, Y. Long, P. Wang, D. Chen, Z. Yang, H. Liang, M. Xue, H. Weng, Z. Fang, X. Dai, and G. Chen, Phys. Rev. X **5**, 031023 (2015).

[19] M. Hoesch, T. Greber, V. N. Petrov, M. Muntwiler, M. Hengsberger, W. Auwaerter, and J. Osterwalder, J. Electron Spectrosc. Relat. Phenom. **124**, 263 (2002).

[20] F. Meier, J. H. Dil, and J. Osterwalder, New J. Phys. **11**, 125008 (2009).

[21] U. Heinzmann and J. H. Dil, J. Phys. Condens. Matter **24**, 173001 (2012).

[22] K. Starke, A. P. Kaduwela, Y. Liu, P. D. Johnson, M. A. van Hove, C. S. Fadley, V. Chakarian, E. E. Chaban, G. Meigs, and C. T. Chen, Phys. Rev. B **53**, R10544 (1996).

[23] C. Jozwiak, Y. L. Chen, A. V. Fedorov, J. G. Analytis, C. R. Rotundu, A. K. Schmid, J. D. Denlinger, Y.-D. Chuang, D.-H. Lee, I. R. Fisher, R. J. Birgeneau, Z.-X. Shen, Z. Hussain, and A. Lanzara, Phys. Rev. B **84**, 165113 (2011).

[24] S. Suga, K. Sakamoto, T. Okuda, K. Miyamoto, K. Kuroda, A. Sekiyama, J. Yamaguchi, H. Fujiwara, A. Irizawa, T. Ito, S. Kimura, T. Balashov, W. Wulfhekel, S. Yeo, F. Iga, and S. Imada, J. Phys. Soc. Jpn. **83**, 014705 (2014).

[25] Z. Wang, Y. Sun, X.-Q. Chen, C. Franchini, G. Xu, H. Weng, X. Dai, and Z. Fang, Phys. Rev. B **85**, 195320 (2012).

[26] Z. Wang, H. Weng, Q. Wu, X. Dai, and Z. Fang, Phys. Rev. B **88**, 125427 (2013).

[27] Y. Sun, S.-C. Wu, and B. Yan, Phys. Rev. B **92**, 115428 (2015).

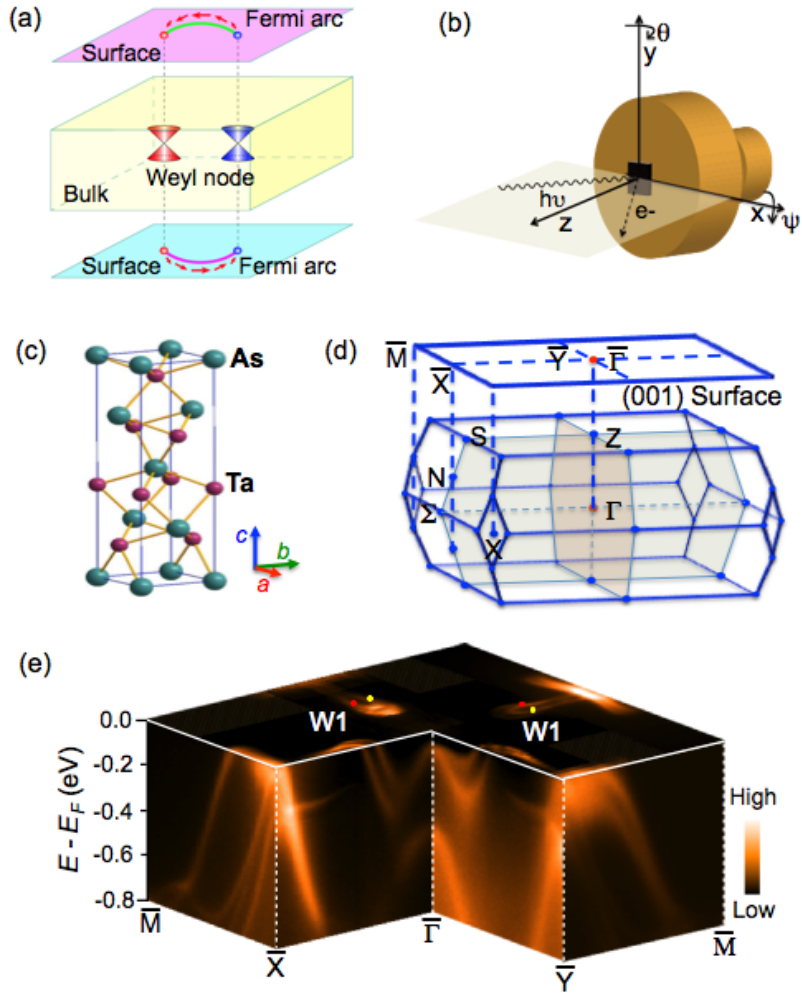

FIG. 1. (Color online). (a) Schematic of a WSM with spin-polarized Fermi arcs on its surfaces connecting the projections of two Weyl nodes with opposite chirality. The red and blue colors of the bulk Weyl cones and the corresponding projection points on the surfaces represent opposite chirality of the Weyl nodes. The red arrows on the surfaces indicate the spin texture of the Fermi arcs. (b) Schematic illustration of the SARPES measurement geometry. The incident beam of light and the outgoing electrons entering the detector define the outgoing plane. The incident light and the hemispherical analyzer form a 45° angle. The reciprocal space along $k_x$ and $k_y$ is explored by varying the angles ψ and θ, respectively. (c) Crystal structure of TaAs. (d) Bulk and projected (001) surface BZs of TaAs. (e) 3D intensity plot of the photoemission data recorded by conventional ARPES measurements, showing the FSs and band dispersions along high-symmetry lines of TaAs. Red and yellow solid circles indicate the locations of the bulk Weyl nodes W1 projected onto the (001) surface BZ, with red and yellow representing opposite chirality.

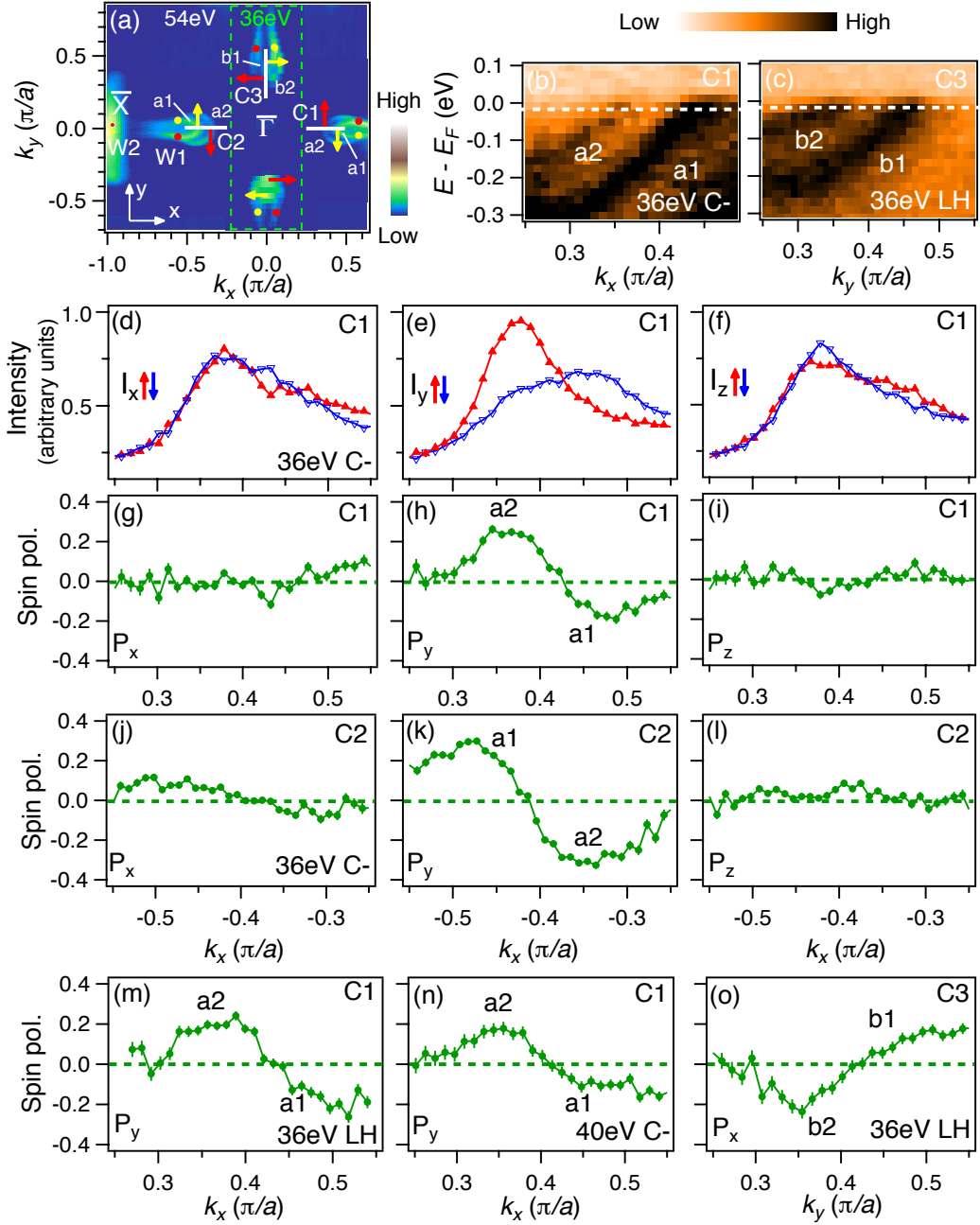

FIG. 2. (Color online). (a) Photoemission intensity plot at $E_F$ of the (001) surface recorded at $h\nu$ = 54 (outside the green box) and 36 eV (inside the green box) recorded by conventional ARPES systems at the Dreamline and SIS beamlines. Red and yellow solid circles indicate the locations of the bulk Weyl nodes projected onto the (001) surface BZ, with red and yellow representing opposite chirality. The red and yellow arrows indicate the direction of spin polarization on the inner and outer FSs at the high-symmetry lines, respectively. The in-plane spin polarization $x$ and $y$ axes are defined along the $\overline{\Gamma}-\overline{X}$ and $\overline{\Gamma}-\overline{Y}$ directions, respectively, and the out-of-plane spin polarization $z$ axis is defined as the outward normal direction of the sample

surface. (b) and (c) Spin-integrated energy–momentum intensity cuts along C1 and C3 [indicated in (a)], respectively. The data were taken with the spin-resolved system at the COPHEE endstation. (d-f) Spin-resolved intensity along C1 projected on the $x$, $y$ and $z$ directions, respectively, measured at $hv = 36$ eV with left-hand circular (C-) polarization. The red and blue symbols are the intensity of spin-up and spin-down states, respectively. (g-i) Corresponding angle-resolved spin polarizations in the $x$, $y$ and $z$ directions, respectively. (j-l) Same as (g-i) but measured along C2. (m) Same as (h) but measured with linear horizontal (LH) polarization. (n) Same as (h) but measured at $hv = 40$ eV. (o) Spin polarization along C3 projected on the $x$ direction measured at $hv = 36$ eV with LH polarization. All the spectra are measured at a binding energy of 20 meV, indicated as white dashed lines in (b) and (c).

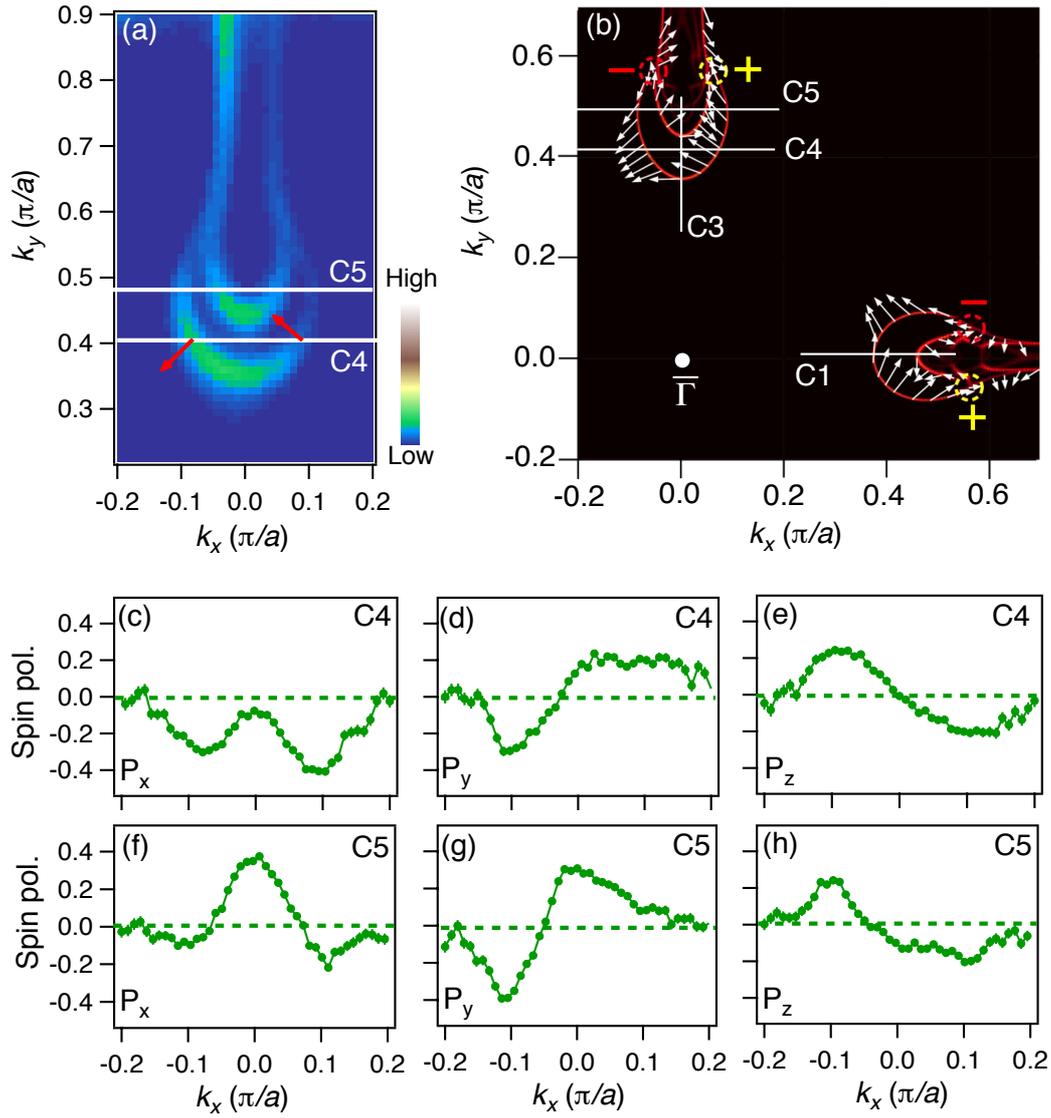

FIG. 3. (Color online). (a) Spin-integrated FS map near $\overline{\Gamma}-\overline{Y}$ recorded with the spin-resolved system at the COPHEE endstation. The red arrows indicate the direction of measured in-plane spin polarizations of the Fermi arc $b2$ at $C4$. (b) Corresponding theoretical spin texture of surface states, with white lines indicating the locations of the SARPES measurements, labeled as $C1$, $C3$, $C4$ and $C5$. Red and yellow dashed circles indicate the Weyl nodes W1 with negative and positive chirality, respectively. (c–e) Angle-resolved spin polarizations along $C4$ in the $x$, $y$ and $z$ directions, respectively. (f-h) Same as (c-e), but along $C5$. All the spectra in Fig. 3 were measured at $hv = 36$ eV with C- polarization and at a binding energy of 20 meV.

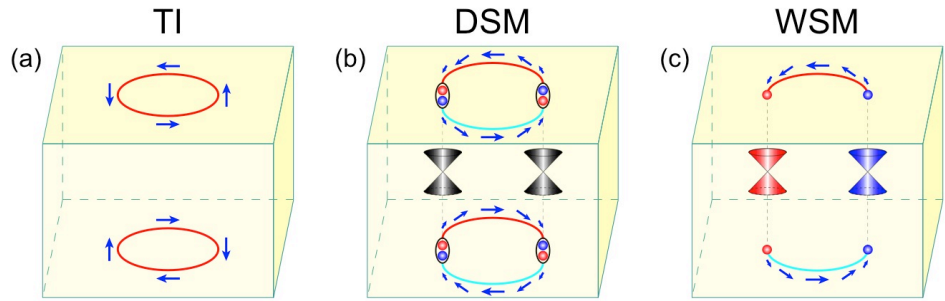

FIG. 4. (Color online). (a) Schematic of the spin-polarized surface states in a 3D TI. The blue arrows indicate the spin polarizations of the surface states. (b) Schematic of a DSM with spin-polarized Fermi arcs on its surface connecting projections of two bulk Dirac nodes. The black color of Dirac cones indicates that one Dirac node is the degeneracy of two Weyl nodes with opposite chirality. (c) Schematic of a WSM with spin-polarized Fermi arcs on its surface connecting projections of two bulk Weyl nodes. The red and blue colors represent opposite chirality. For clarity, only surface states on top and bottom surfaces are indicated.